\documentclass[fleqn,10pt]{wlscirep}
\usepackage[utf8]{inputenc}
\usepackage[T1]{fontenc}
\usepackage{graphicx}
\usepackage{float}	
\usepackage{ucs}
\usepackage{amsmath}  
\usepackage{amsfonts}
\usepackage{amssymb}
\usepackage{booktabs}
\usepackage{textcomp}
\usepackage{gensymb}
\usepackage{enumitem}
\usepackage{multirow}

\title{Revealing time’s secrets at the National Theatre of Costa Rica: innovative software for cultural heritage research}

\author[1,2]{M.D. Barrantes--Madrigal}
\author[3]{T. Zúñiga--Salas}
\author[4]{R.E. Arce--Tucker}
\author[2,5]{A. Chavarría--Sibaja}
\author[6]{J. Sánchez--Solís}
\author[5]{J. Mena--Vega}
\author[2,5]{K. Acuña--Umaña}
\author[1]{M. Gómez--Tencio}
\author[3]{K. Wang--Qiu}
\author[1]{F. Lizano--Fallas}
\author[7]{C. Marín--Cruz}
\author[2,5,8,9,*]{O.A. Herrera--Sancho}

\affil[1]{Escuela de Química, Universidad de Costa Rica, 2060 San Pedro, San José, Costa Rica}
\affil[2]{Centro de Investigación en Ciencias Atómicas Nucleares y Moleculares, Universidad de Costa Rica, 2060 San Pedro, San José, Costa Rica}
\affil[3]{Escuela de Artes Plásticas, Universidad de Costa Rica, 2060 San Pedro, San José, Costa Rica}
\affil[4]{Facultad de Microbiología, Universidad de Costa Rica, 2060 San Pedro, San José, Costa Rica}
\affil[5]{Escuela de Física, Universidad de Costa Rica, 2060 San Pedro, San José, Costa Rica}
\affil[6]{Escuela de Ingeniería Eléctrica, Universidad de Costa Rica, 2060 San Pedro, San José, Costa Rica}
\affil[7]{Teatro Nacional de Costa Rica, 5015--1000 San José, Costa Rica}
\affil[8]{Centro de Investigación en Ciencia e Ingeniería de Materiales, Universidad de Costa Rica, 2060 San Pedro, San José, Costa Rica}
\affil[9]{Instituto de Investigaciones en Arte, Universidad de Costa Rica, 2060 San Pedro, San José, Costa Rica}

\affil[*]{oscar.herrerasancho@ucr.ac.cr}
\begin{abstract}

Establishing affordable, efficient, accessible, innovative, and multidisciplinary methodologies is key to the creation of modern public policies on cultural heritage. Limited access to large--format paintings is a challenge to restoration scientists seeking to obtain information quickly, in a non--destructive and non--invasive manner, and identify regions of interest. Therefore, we put forward two unique software tools based on multispectral imaging techniques, with the long--term aim of assessing the artist's intentions, creative process, and colour palette. This development paves the way for a comprehensive and multidisciplinary understanding of the mysteries encompassed in each pictorial layer, through the study of their physical and chemical characteristics. We conducted the first ever study on \textit{Musas I} and \textit{Musas II}, two large--format paintings by Italian artist Carlo Ferrario, located in the National Theatre of Costa Rica. In this study, we used our novel imaging techniques to chose regions of interest in order to study sample layers; while also assessing the works' state of conservation and possible biodeterioration. We explored the applications of our two versatile software tools, \textit{RegionOfInterest} and \textit{CrystalDistribution}, and confirmed paint stratigraphies by means of microscopy and spectroscopy analyses (OM, SEM--EDX, Fluorescent microscopy, FTIR--ATR and micro--Raman). In a pilot study, we identified the artist’s main colour palette: zinc white, lead white, chrome yellow, lead read, viridian, along with artificial vermilion and ultramarine pigments. We were able to identify artificial vermilion and ultramarine and distinguish them from the natural pigments using \textit{CrystalDistribution} to map the average size and diameter of the pigment crystals within the paint layers. This study demonstrated that software--based multidisciplinary imaging techniques are fundamental in establishing novel preventive and non--invasive methods for historical painting conservation studies and virtual restoration.

\end{abstract}

\begin{document}

\flushbottom
\maketitle
\thispagestyle{empty}

\section*{Introduction}

Art opens a window into the historical context in which it was created. Studying the paintings of a population offers insight into its culture, history and origins. In order to protect these expressions of individual and collective identity, scientists have set about conserving and restoring works of art. In the case of large--format paintings, today's conservation efforts demand affordable, effective, accessible, innovative, and multidisciplinary methodologies~\cite{fiorillo2020, madariaga2015}. Typically, conservation studies involve chromatography, microscopy, spectroscopy and electrochemical techniques~\cite{madariaga2015, ortiz2017development}, alongside biological analyses to identify: material composition~\cite{artesani2019}, painting technique, and microscopic damage~\cite{fiorillo2020, ortiz2017development}. Microscopy and spectroscopy techniques have been widely used to examine the stratigraphy and composition of materials~\cite{madariaga2015, artesani2019} in small painting samples~\cite{mahmoud_2014}. These techniques are efficient and accurate, but the equipment is very costly~\cite{madariaga2015}, opportunities for sample acquisition are limited, and results obtained for specific samples cannot be applied to the painting as a whole in large--format works~\cite{delaney2016}. Limited access represents a further drawback, as restoration scientists are challenged to obtain sufficient information quickly, in a non--destructive, and non--invasive manner, when identifying regions of interest.

Optical imaging is a major area of development in artwork diagnostics. Novel imaging tools such as photoacoustic imaging~\cite{Tserevelakis12017} and CT--scanning~\cite{sallam2019}
have recently been used to study otherwise inaccessible features of art objects. In this study we used Multispectral Imaging (MSI), an affordable, non--invasive and non--destructive technique which relies on a photography system to map and identify pigments, binders, and retouched areas of the paintings~\cite{cosentino2015}. MSI allowed us to verify the painting technique and evaluate the current state of the paintings, including damage and possible restorations~\cite{fiorillo2020, liang2011}. In order to quantitatively examine the colours, most studies have recreated reflectography spectra for a selected painting region using MSI; however, the main drawback of this technique is that it requires many filters and it is very time--consuming~\cite{cosentino2015, liang2011}. Nowadays, MSI can be coupled with computational tools, although in most cases, this is used to simulate restoration~\cite{barni2005} and improve the image quality~\cite{ribes2008, paviotti2009lightness}. Nonetheless, there is still an evident need for new quantitative MSI techniques to assess the artist's intentions, creative process, and colour palette used in the artworks. 

This paper presents two novel computational tools: \textit{RegionOfInterest} and \textit{CrystalDistribution}, which we believe to be versatile and accessible elements for non--invasive conservation studies worldwide, particularly in the case of historical paintings and virtual restoration. With these advanced computational tools we are able to gather otherwise hidden information on works of art; such as the number and size of pigment crystals in a given area. This analysis reveals specific characteristics of the pigments in a work; for example, if they are of natural or artificial origin. Using these cutting--edge multispectral imaging techniques, we are also able to obtain quantitative luminosity data from MSI. The \textit{RegionOfInterest} program can accurately calculate luminosity intensity within user--selected regions in a photo of the painting, and can outline all the areas of the painting that present a given intensity value. This should prove to be a useful tool for artistic and heritage studies with both micro and macroscopic approaches, since it can identify the pigments used in a simple sample and then identify all regions of interest in the artwork where that pigment may be present. More specifically, we believe these tools represent an important step forward in restoration efforts and preventive methods, since they allow for guided microbiological sampling, even before fungi have caused visible damage. Using these computational tools in a pilot study, regions of interest were isolated from the rest of the artwork in order to study stratigraphies of individual samples and carry out microbiological sampling, alongside environmental monitoring of the painting’s surroundings, as these are key factors for efficient modern public policies in cultural heritage~\cite{fiorillo2020}.

\begin{figure}[ht]
    \centering
     \includegraphics[scale=0.1]{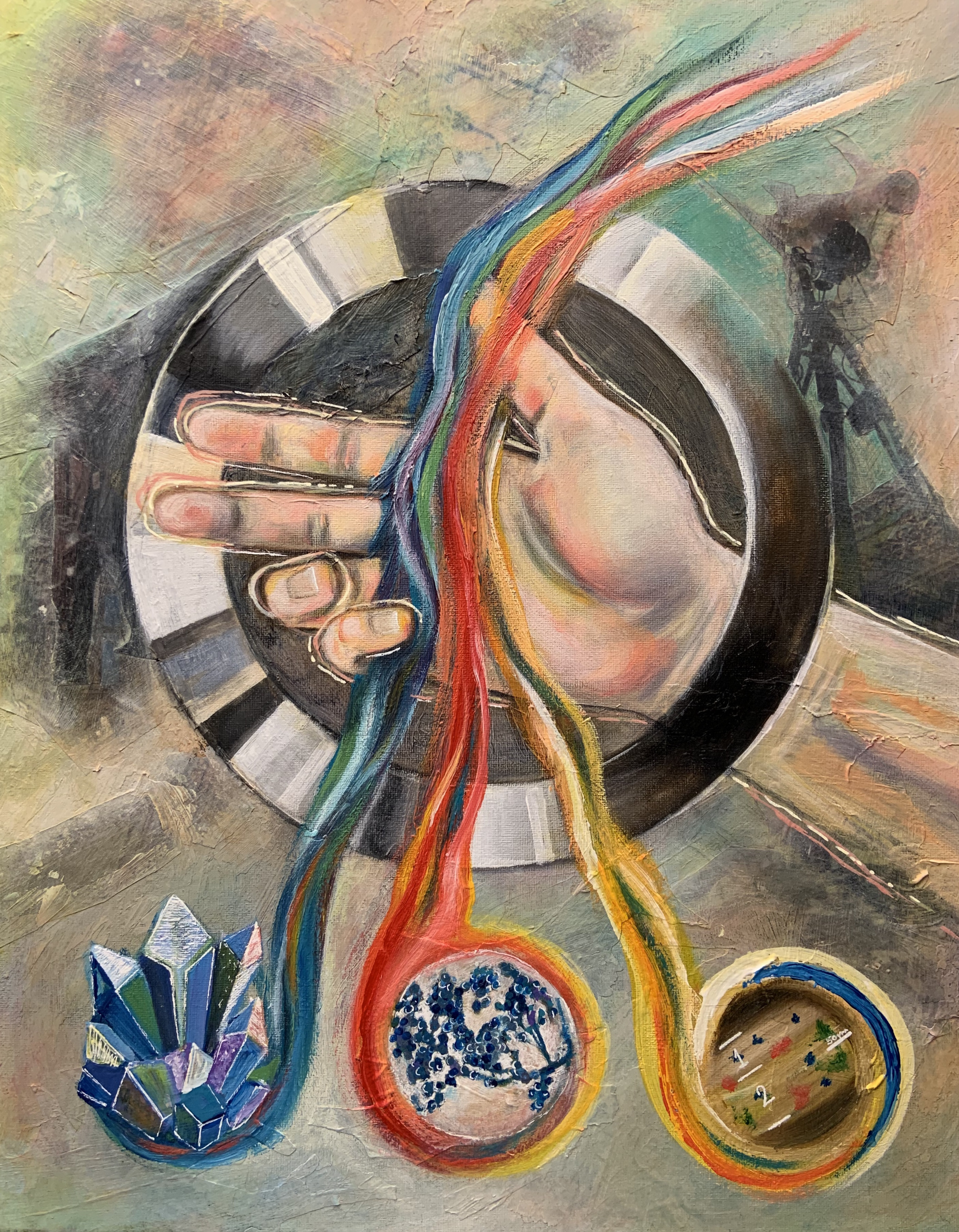}
     \caption{\textbf{In the flow of time}. This original painting tells the story of the study, describing our journey as we delve into the hidden secrets behind large--format artworks. A fluctuating and changing process, our multi--analytical study's use of novel software tools and multispectral imaging allowed us to fantasize, even for a second, that we could stop the passage of time and capture the artist's creative process, both materially and conceptually. As the colours flow through our fingers, we perceive individual pigment crystals, sense the damage caused by the passage of time, understand the materials and layers that come together to form the painting, and almost hear the artist's voice as we unveil his intentions.
     Although created using contemporary techniques such as acrylic and collage on canvas, this piece was inspired by Carlo Ferrario's creative process. His colour palette influenced our own, as we selected pigments found in \textit{Musas I} and \textit{Musas II} such as ultramarine blue, vermilion, and white. Our painting is structured similarly to Ferrario's, with two layers of ground, two layers of paint, and a final layer of varnish.
    Art, like history and time, slips inexorably through our fingers. Today, we capture in this study an instant of that flow, gaining new perspectives and revealing some of the hidden secrets behind large-format paintings.}
    \label{figure1}
\end{figure}

Art conservation is a complex process that has traditionally relied on both micro and macroscopic techniques, involving a comprehensive and multidisciplinary understanding of a work's physical and chemical characteristics. Figure~\ref{figure1} outlines our vision of how to systematically tackle the complexity of large--format works. This investigation approaches large--format paintings through a multi--analytical study (MAS) which includes MSI analyses, microscopic, spectroscopic, microclimate and microbiological analyses, coupled with novel computational tools. Thus, the main objective of the present study is to establish a baseline diagnostic of the state of conservation of paintings \textit{Musas I} and \textit{Musas II} by Italian painter Carlo Ferrario, located in the National Theatre of Costa Rica (NTCR), using advanced computational imaging techniques to carefully choose regions of interest.

A multidisciplinary approach has been used to great effect on previous art conservation studies. For instance, a study in Linares, Spain combined the use of Energy Dispersive X--ray Fluorescence ($\mu$EDXRF), surface mapping and single--spot micro-Raman spectroscopy to identify the chemical composition of pigment layers in wall paintings at the Cástulo site~\cite{tunon2020}. Another MAS was carried out on the large--format painting \textit{Marriage at Cana} by Italian artist Luca Longh, located in Ravenna, Italy, to identify its technique and state of conservation~\cite{fiorillo2020}. This study revealed the materials used in the work, and that it was executed using the dry painting technique. This MAS shed light on how restoration had been conducted, and biological analyses identified the \textit{Eurotium halophilicum} fungus damaging the painting. This information was vital to the cleaning and restoration processes. Our study has similar objectives: to contribute significantly to the conservation of the paintings \textit{Musas I} and \textit{Musas II}, by another Italian painter, Carlo Ferrario; the essential difference being that target regions were precisely knitted within the artworks which are located in Central America.

\subsection*{Historical context}

The construction of the National Theatre of Costa Rica (NTCR) (begun in 1891 and completed in 1897) stood as a symbol of the growing coffee oligarchy in Costa Rica~\cite{Marin2012Metodologia}, for its building, were hired engineers, architects, painters, sculptors, and decorators from Europe, particularly from Italy \cite{Fischel1981caja}. The subjects of this study are \textit{Musas I} (2.96\,m (w), 6.17\,m (h)) and \textit{Musas II} (2.96\,m (w), 6.17\,m (h)), two large-format paintings by italian artist Carlo Ferrario, emplaced on the ceiling roughly 3.5\,m above the ground and originally located in the ladies' lounge inside the theatre~\cite{Fischel1981caja}, see Fig.~\ref{figure2}a (top) and Fig.~\ref{figure2}a (bottom). Born in Milan in 1833, Ferrario had extensive experience as a designer, decorator, and painter of scenery and curtains in prestigious Italian theatres. Although he painted \textit{Musas I} and \textit{Musas II} in his workshop in Italy~\cite{Fischel1981caja}, he came to Costa Rica to assist in the decoration of the NTCR and settled in once he finished the work at the NTCR~\cite{Fischel1981caja}. There are reports that \textit{Musas I} and \textit{Musas II} have been previously restored, however there are no descriptions or documentation of the type of work performed. In previous studies of artwork at the NTCR, we were able to examine only a limited number of samples, without first identifying areas of interest~\cite{Morice2019, Conejo2020}. This is the first time we have had the opportunity to take a comprehensive, multidisciplinary approach.

Tropical countries raise different challenges about the conservation and restoration of paintings. Art pieces imported into Latin America were not designed to withstand the effects of the temperatures and humidity of the tropics. There are many studies on the conservation and restoration of large--format artworks, involving the techniques discussed above, but they took place in Europe \cite{fiorillo2020, marras2010study, petrova2019pigment, paradisi2012domus, gil2020old} and did not address tropical factors. There are even fewer restoration protocols suited to the works in question. Deterioration caused by tropical conditions is still largely unstudied, underscoring the importance of the present investigation.

This article describes the study of paintings \textit{Musas I} and \textit{Musas II} through a multi--analysis approach that includes: (1) MSI coupled with novel computational tools for image analysis, (2) microscopy and spectroscopy analyses with the development of custom software to identify the size of pigment crystals observed in stratigraphies of the samples, (3) microbiological analysis for fungi identification in previously defined regions of interest, and (4) monitoring of environmental conditions in the paintings’ surroundings, in order to present a general diagnostic of the paintings.

\section*{Results}

Carlo Ferrario painted \textit{Musas I} and \textit{Musas II} 123 years ago. Today, we contemplate our heritage and wonder: what history is woven into these paintings? What secrets lurk in each brushstroke, pigment, fiber and crack in the canvas? Why did the artist choose each material? Figure \ref{figure1} symbolizes our multi--disciplinary approach to these questions, as we draw on diverse fields of study, tools and techniques to better understand the paintings' history and assess their current state of conservation. It is important to note that restoring large--format paintings does not seek to turn back time or erase history~\cite{brandi2005}, but rather to preserve it for future generations. 

\subsection*{More than meets the eye: Multispectral imaging}

\begin{figure}
    \centering
    \includegraphics[scale=0.35]{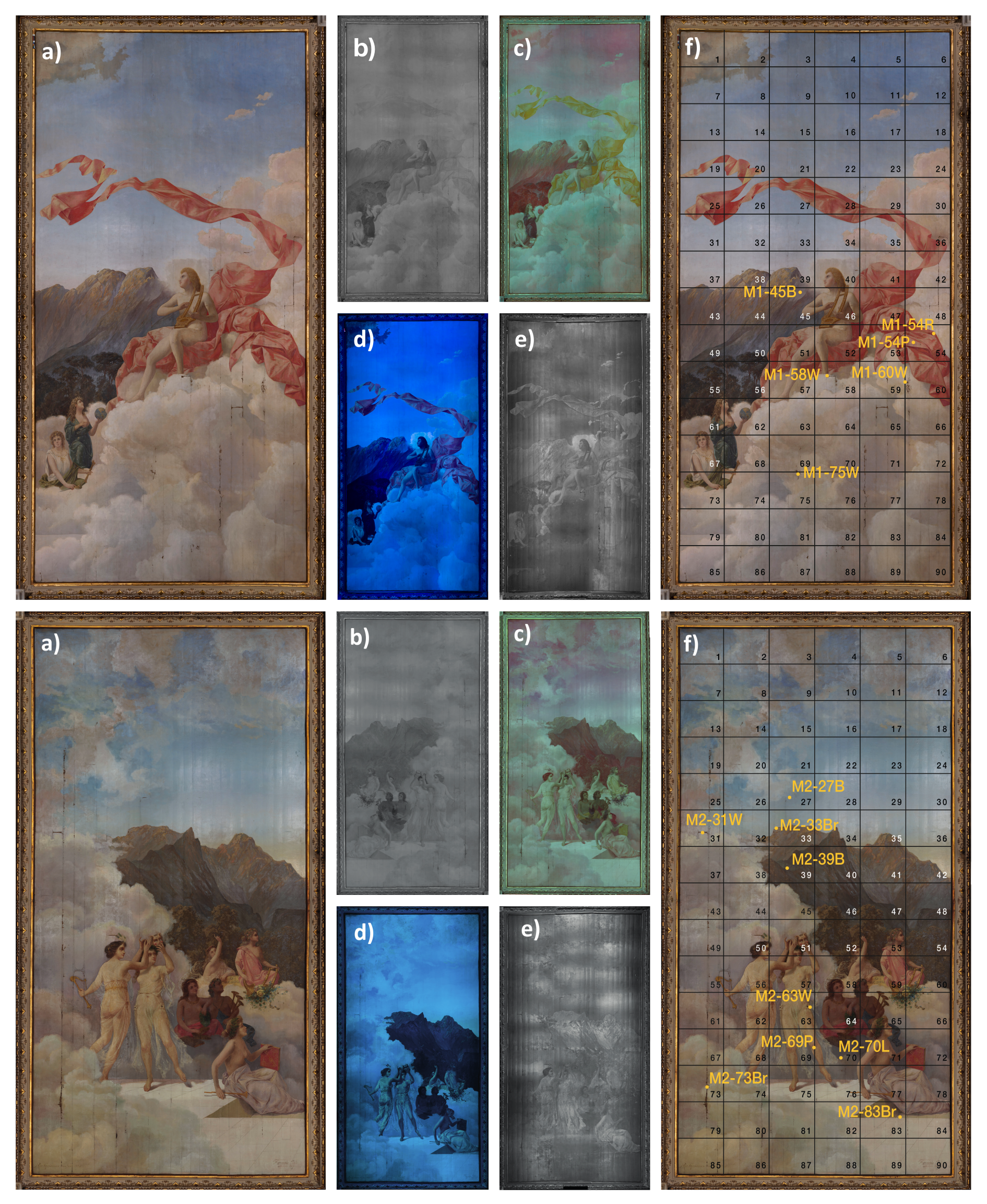}
    \caption{\textbf{Six views at once: Multispectral Imaging of \textit{Musas I} (top panel) and \textit{Musas II} (bottom panel)}, \textbf{(a)} Visible (VIS), \textbf{(b)} infrared (IR), \textbf{(c)} infrared false colour (IRFC), \textbf{(d)} ultraviolet fluorescence (UVF) and \textbf{(e)} ultraviolet reflectance (UVR) multispectral images of \textit{Musas I} and \textit{Musas II} and \textbf{(f)} sample locations on the grid. The M1-XXY codification of sample names indicates: M1 = Painting \textit{Musas I}, M2 = Painting \textit{Musas II}, XX = location on the grid, Y = visible colour of the sample (B = blue, R = red, P = pink, W = white).}
    \label{figure2}
\end{figure}

Multispectral imaging (MSI) reveals information difficult or impossible to see with the naked eye, especially useful with large--format paintings, where exhaustive in--person examination may not be practical. Figure~\ref{figure2} (top) and Figure~\ref{figure2} (bottom) show multispectral images of \textit{Musas I} and \textit{Musas II}, respectively. Five spectral regions are shown: visible (VIS) in Fig.~\ref{figure2}a (top) and Fig.~\ref{figure2}a (bottom), infrared (IR) in Fig.~\ref{figure2}b (top) and Fig.~\ref{figure2}b (bottom), infrared false colour (IRFC) in Fig.~\ref{figure2}c (top) and Fig.~\ref{figure2}c (bottom), ultraviolet fluorescence (UVF) in Fig.~\ref{figure2}d (top) and Fig.~\ref{figure2}d (bottom), and ultraviolet reflected (UVR) in Fig.~\ref{figure2}e (top) and Fig.~\ref{figure2}e (bottom).
VIS images show that \textit{Musas I} and \textit{Musas II} are complex systems of detailed artistry, including myriad colours, glazes, textures and brushstrokes, some of which may be a product of restoration efforts. The materials interact with one another and with environmental factors over time, as the paintings have been mounted on a wooden ceiling for more than 120 years. We can observe some damage in Fig.~\ref{figure2}a (top) and Fig.~\ref{figure2}a (bottom). Infrared images in Fig.~\ref{figure2}b (top) and Fig.~\ref{figure2}b (bottom) show inner layers of the paintings, because some external layers do not absorb infrared radiation, rendering them transparent~\cite{fiorillo2020}. Other materials appear dark, such as the dress of the woman holding the Earth in \textit{Musas I}, see Fig.~\ref{figure2}b. Pigments in this area absorb infrared radiation~\cite{cosentino2014identification}, while the reflective pigments opposite them look brighter~\cite{cosentino2014identification}. IRFC images combine channels from both VIS and infrared images, revealing particle pigment behaviours.

UVF shows the fluorescence of each material. Varnish usually fluoresces strongly, overshadowing pigment fluorescence~\cite{cosentino2015practical}. This effect is not observed in \textit{Musas I} and \textit{Musas II} (Fig.~\ref{figure2}d (top) and Fig.~\ref{figure2}d (bottom)) because the varnish was removed with ammonia in 1997, according to NTCR reports. The UVF pictures also show: the exposed canvas; brighter areas of the clouds may indicate the use of different binders; stains caused by insects or microorganisms; and possible areas of restoration, such as the white cloud behind the head of the main muse on Fig.~\ref{figure2}d (top). Meanwhile, UVR (Fig.~\ref{figure2}e (top) and Fig.~\ref{figure2}e (bottom)) provides information on the outermost layers~\cite{cosentino2014identification} and biological staining. UVR was used primarily to identify white pigments~\cite{fiorillo2020,madariaga2015}. Titanium and zinc white appear dark, while lead and lithopone appear bright~\cite{madariaga2015}, so we can infer that the bright regions in Fig.~\ref{figure2}e (top) and Fig.~\ref{figure2}e (bottom) correspond to lead white or lithopone.

\subsection*{Deeper secrets in heritage research: developing software for closer analysis}

Further MSI analyses are needed to reveal underlying layers, sketches and corrections, possible previous restoration and other regions of interest. We believe software based on multispectral imaging can offer a new perspective in the study of art. We developed \textit{RegionsOfInterest} to classify colour luminosity, showing chromatic distributions in large--format works. This is relevant to art history, art restoration, and even an artist's creative process.

\begin{figure}
    \centering
    \includegraphics[scale=0.36, angle=270,origin=c]{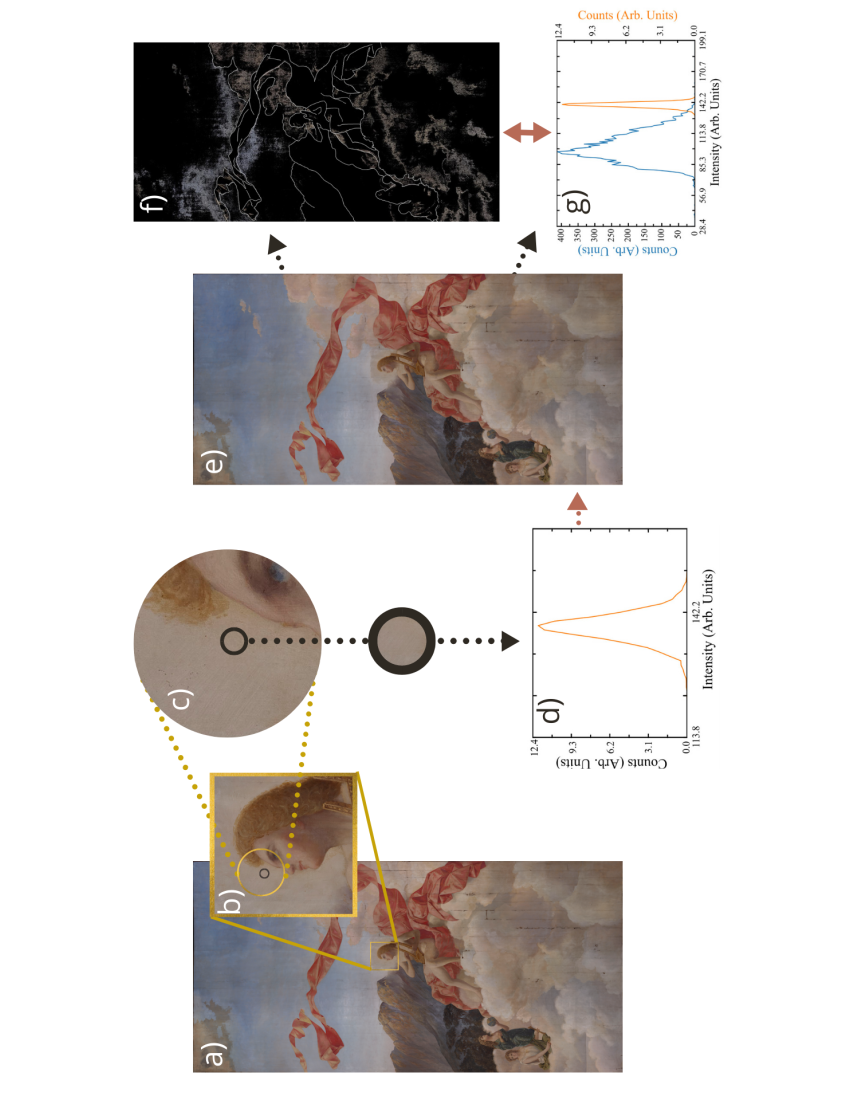}
    \caption{\textbf{A closer look: Execution of \textit{RegionOfInterest} software.} \textbf{(a)} The program opens the image for general user inspection. \textbf{(b)} In the first stage, the user zooms in on a region of interest. \textbf{(c)} The user hand--draws a contour to select the area for analysis. \textbf{(d)} The program converts the selected pixels from HSV to Intensity space and automatically generates a "Counts vs Intensity" histogram indicating pixel intensity distribution ranging from 0 to 255. \textbf{(e)} In the second stage, the user chooses the range from the first histogram. \textbf{(f)} Pixels in the full image that fall within this intensity range are displayed in a new colour image. \textbf{(g)} A second intensity histogram is generated, comparing the intensity distribution of all pixels in the original image (blue line) with that of the selected pixels in the new image (orange line). Note that the y--axis shows different scales for the blue and orange curves, left and right, respectively.}
    \label{figure3}
\end{figure}

Figure~\ref{figure3} illustrates the \textit{RegionsOfInterest} tool, developed specially for restoration scientists:

\begin{itemize}
    \item The user uploads an image of the artwork in .jpg, .png, etc, see Fig.~\ref{figure3}a.
    \item The user selects a sample area. The colours from this area will be mapped to the artwork as a whole, in order to determine the spatial distribution of areas of interest such as colour, pigment and other zones (see Fig.~\ref{figure3}b and Fig.~\ref{figure3}c).
    \item The histogram in Fig.~\ref{figure3}d shows this region's intensity values.
    \item \textit{RegionsOfInterest} displays regions in the specified intensity range on the artwork as a whole, see Fig.~\ref{figure3}f. 
    \item Lastly, Fig.~\ref{figure3}g shows an increase in the number of pixel intensity counts for this region.        
\end{itemize}

\begin{figure}
    \centering
    \includegraphics[scale=0.36, angle=270,origin=c]{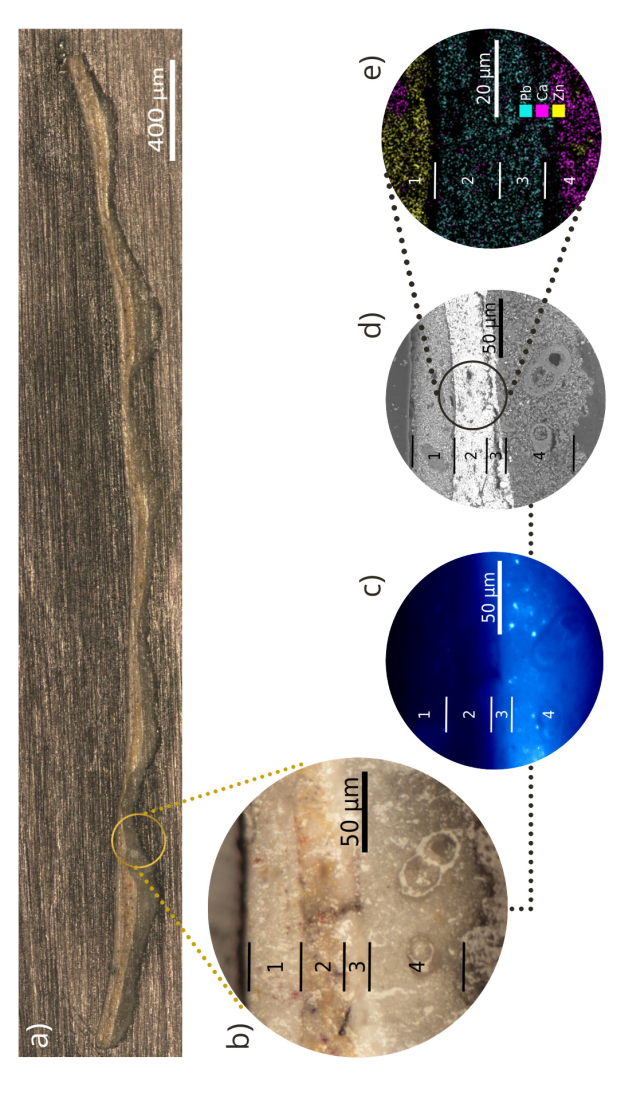}
    \caption{\textbf{Through the lens: Stratigraphy of a 120--year--old large--format painting.} \textbf{(a)} Panoramic Optical Microscopy (OM) image of the cross--sectioned sample M1--75W, observed by reflected visible light at 5X. \textbf{(b)} Detail of the sample observed by MO at 20X showing the main layers: (1) Top paint layer, (2) Intermediate paint layer, (3) Second ground layer, (4) First ground layer. \textbf{(c)} Fluorescence Microscopy showing fluorescent particles in layer 4. \textbf{(d)} Scanning Electron Microscopy (SEM) showing the area analyzed by SEM--EDX using the mapping tool. \textbf{(e)} The main elements in each layer.}
\label{figure4}
\end{figure}

\begin{figure} 
    \centering
    \includegraphics[scale=0.07]{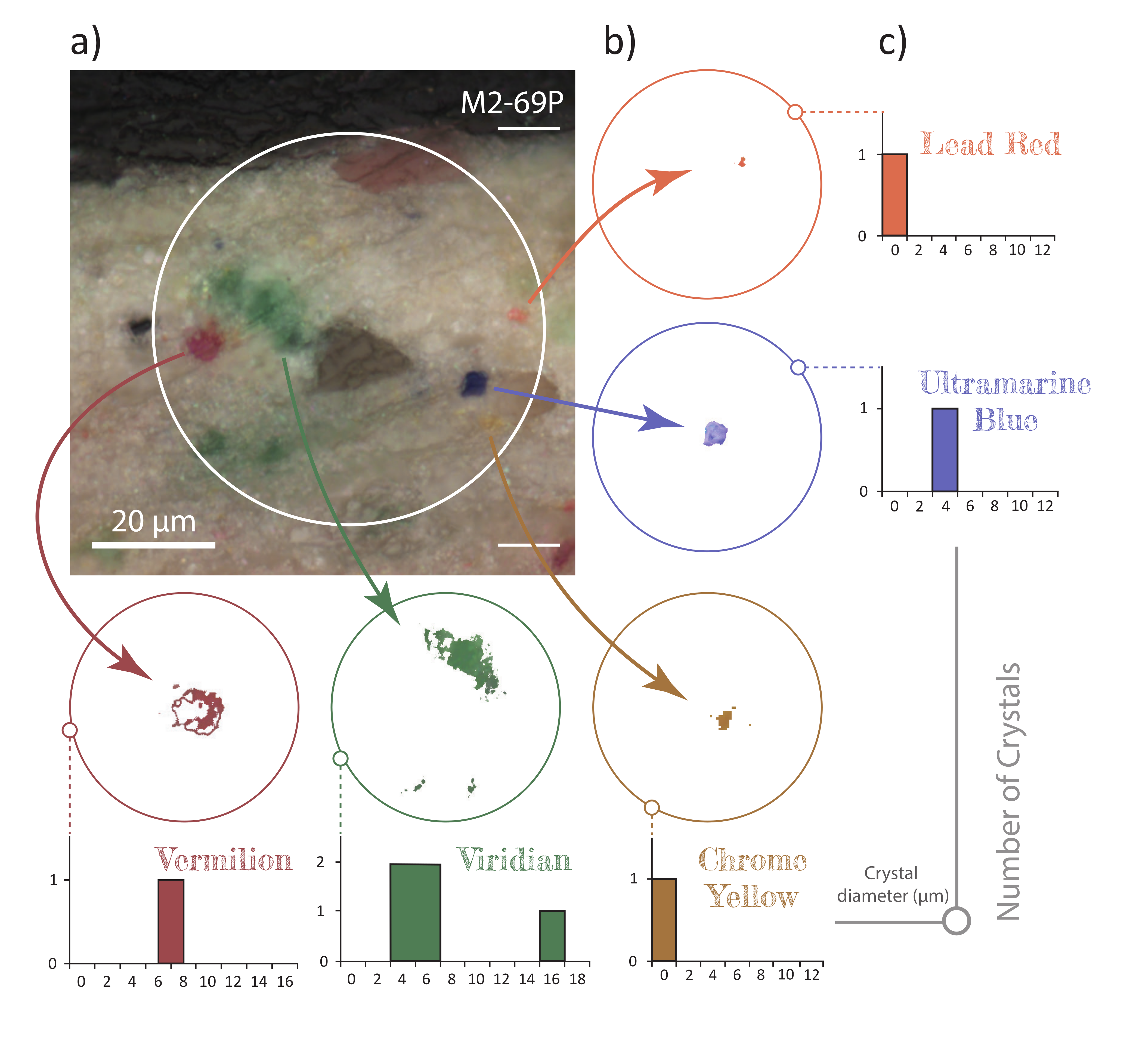}
    \caption{\textbf{Between particles: Measuring crystals with \textit{CrystalDistribution} software} \textbf{(a)} The user opens a microscope image and select the region containing crystals of interest. \textbf{(b)} The program applies masks or "cuts" identifying the crystals and saves the images. Each image shows the crystals of a specific colour. \textbf{(c)} The program calculates the crystals' area and diameter, and generates a histogram for each colour.}
    \label{figure5}
\end{figure}

\begin{figure}
    \centering
    \includegraphics[scale=0.4]{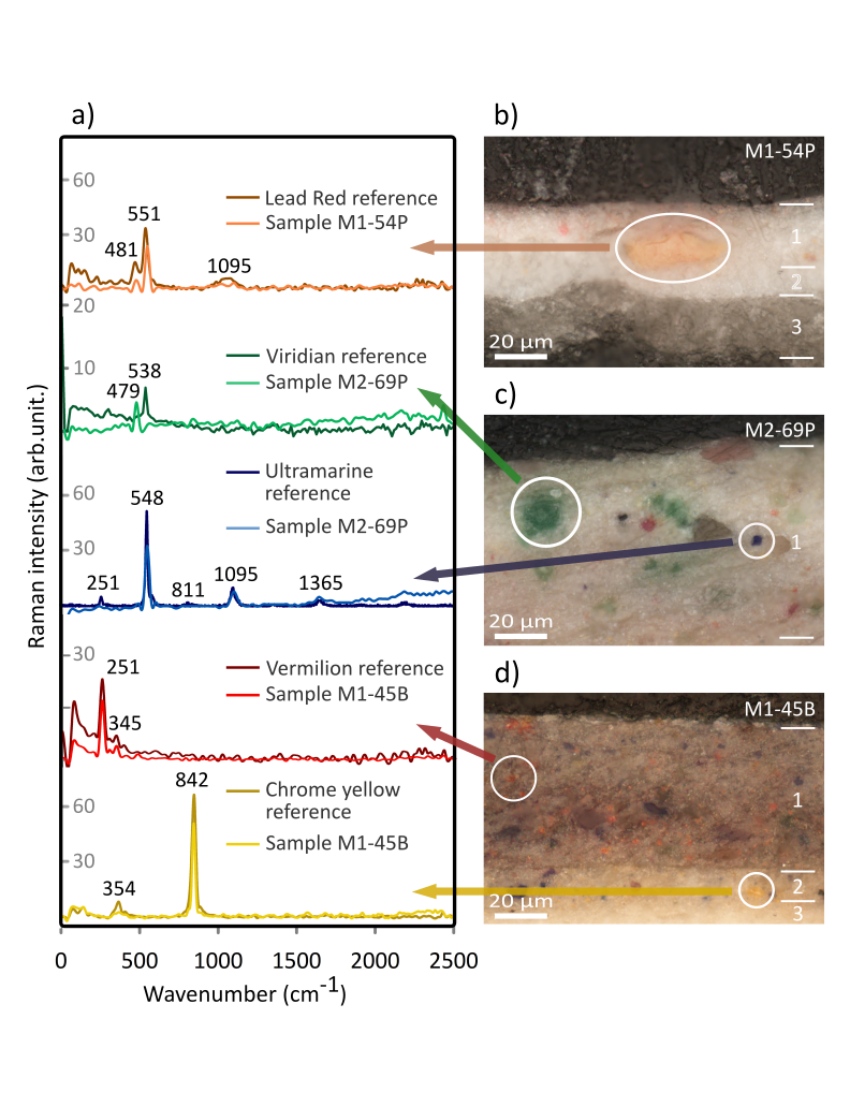}
    \caption{\textbf{In low frequency: Raman spectroscopy of the pigments}. \textbf{(a)} Comparison of Raman spectra for reference pigments and those of the samples \textbf{(b)} Orangey--pink crystal from sample M1--54P corresponds to lead red, \textbf{(c)} Green and blue crystals from sample M2--69P correspond to viridian and ultramarine, respectively. \textbf{(d)} Red and yellow crystals from sample M1--45B correspond to vermilion and chrome yellow, respectively.}
    \label{figure6}
\end{figure}

The software can be used to systematically select regions of interest for comprehensive and efficient non--invasive sampling, for conservation diagnostics. This novel and time--saving tool has three applications:

\begin{enumerate}

    \item \subsubsection*{Comparing colour intensity behaviours} 

Firstly, we characterized a range of whites, which appear to be the most abundant colour in the artwork. Analysis of 11 regions in \textit{Musas I} using \textit{RegionsOfInterest} resulted in a pixel intensity value of approximately 132\,$\pm$12. Since similar values are obtained for white regions containing similar compounds, intensity analysis software is a useful tool during the composition stage of the creative process. For instance, the artist can establish a more complete and complex colour palette based on colour saturation by analyzing colour intensity distribution with \textit{RegionsOfInterest}.

    \item \subsubsection*{Revealing the colour palette} 

Secondly, following visual analysis of the painting and colour palettes prevalent at the time, we hypothesized that all colours in the paintings were obtained by mixing approximately 12 original pigments. As a proof--of--principle, we analyzed the paintings with the \textit{RegionsOfInterest} program to identify colour distributions that could provide insight into the artist's palette. We examined six colour intensities (yellow, red, blue, green, brown, and white). On average, yellow was the most prevalent, at about 20\% abundance, while green was the least, at only 1\%. The single most striking observation was that these colour intensities make up 50\% of \textit{Musas I} and 30\% of \textit{Musas II}. This application could be useful in other case--studies, to identify an artist’s main palette and the agents responsible for deterioration.
    
Figure~\ref{figure4} illustrates stratigraphies of the sample M1-75W (for sample information, please refer to Table~\ref{table1}), collected from \textit{Musas I} using our software to identify regions of interest, and analyzed through microscopy and spectroscopy (OM, SEM--EDX, and fluorescent microscopy). A total of 15 samples analyzed (6 from \textit{Musas I} and 9 from \textit{Musas II}) displayed similar structures made of up of four layers~\cite{stoner2013conservation}, see Fig.~\ref{figure4}b,  which we identified as: (1) top paint layer, (2) intermediate paint layer, (3) second ground layer, and (4) first ground. We further analyzed the first ground using FTIR-ATR and identified distinctive bands of calcite from calcium carbonate: C--O symmetric stretching (at 2523\,cm$^{-1}$, 1802\,cm$^{-1}$), C--O asymmetric bending (at 1627\,cm$^{-1}$), C--O asymmetric stretching (at 1392\,cm$^{-1}$), C--O symmetric stretching (at 1090\,cm$^{-1}$), C--O asymmetric bending and C--O symmetric bending (at 712\,cm$^{-1}$)~\cite{xu2014linking}. This compound was widely used as a ground material during the nineteenth century~\cite{stoner2013conservation}. We were also interested in a unique morphology of foraminifera: prehistoric marine invertebrates which turned out to be an important source of calcium carbonate, CaCO$_3$~\cite{stoner2013conservation}. Additionally, Fig.~\ref{figure4}c shows fluorescent particles observed through inverted fluorescence microscopy, discussed in the forthcoming section. The elemental composition of each layer identified by SEM--EDX is shown in Fig.~\ref{figure4}d and Fig.~\ref{figure4}e, respectively. We found that, generally, the top paint layer contains an important amount of zinc; layers 2 and 3 contain mostly lead; and the first ground consists of calcium from calcium carbonate, as expected. As we found about 44$\%$ of zinc in some regions of the ground layer, we hypothesize that the ground contains zinc white pigment mixed with calcium carbonate.

To the best of our knowledge, there has been no systematic study of the colour palette used by Carlo Ferrario. In order to contextualize his creative process, we studied the pigments from both the pictorial and ground layers. The density of crystals in an area of approximately 2000\,\,$\mu$m$^{2}$ was arbitrarily categorized as follows: fewer than 4 crystals was considered low density, 4--20 crystals medium density, and more than 20 crystals high density. The average diameter of the crystals was calculated using our custom \textit{CrystalDistribution} software, see Fig.~\ref{figure5}, which shows how each colour was analysed individually to determine crystal size and quantity. We employed micro--Raman spectroscopy to identify the chemical composition of these pigments, see Fig.~\ref{figure6}a. The measured spectra were compared to reference pigments from the Cultural Heritage Open Source Pigments Checker (CHSOS). 

\textit{Lead red}

Lead red pigment in \textit{Musas I} was likely used in areas that appear light red, such as the cloak and other reddish--pink clothing and ornaments. 
Microscopically, large light red with some orange hue crystals were found in the paint layer of sample M1--54P which is located in one of the lighters sections of the mantle, see Figure~\ref{figure2}g (top). The cross--section of this sample shows a low density of crystals with an average diameter of about 21\,$\mu$m, see Fig.~\ref{figure6}b. Lead red, applied over other base colours, such as vermilion, has been commonly used since the middle--ages to give the effect of silk cloth~\cite{feller1986artist}. Ferrario may well have used this technique.
Figure~\ref{figure6}a presents Raman spectra of this crystal, showing characteristic components of lead red, in particular the Pb--O vibrational modes~\cite{edwards1999minium} at 550 cm$^{-1}$. 

\textit{Viridian}

Viridian pigment could be found in green tones in the mountains, background and forest area. Interestingly, among all the samples under study, only M2--69P shows green crystals in its paint stratigraphy, see Fig.~\ref{figure6}c. Low density of these crystals was observed in the paint layer, with an average diameter of roughly 5.1\,$\mu$m, see Fig.~\ref{figure5}c.
Michelangelo used an underlayer of green earth and viridian to create the effect of light~\cite{cardeira2016analytical}. Ferrario may have used the related verdaccio method, in which a neutral colour, usually green, is applied in underpainting for outlining and shading~\cite{mayer1991artist}.
Raman analyses identified the green crystals as viridian, with a reference signal at 538 cm$^{-1}$ for chrome III oxide, Cr$_{2}$O$_{3}$~\cite{maslar2001situ}, see Fig.~\ref{figure6}a. However, the samples' spectra showed a signal at 479 cm$^{-1}$ generated by the Cr$^{(VI)}$-O stretching mode~\cite{brown1968infrared} of dihydrated chrome III oxide Cr$_{2}$O$_{3}$·2H$_{2}$O, the major component of viridian. 
Developed in 1838~\cite{ball2008invention}, this pigment was very expensive until a French chemist devised an alternate production method in 1859, making it more accessible to contemporary artists~\cite{ball2008invention}.

\textit{Ultramarine}

Blue pigment is most evident in the sky areas of both \textit{Musas I} and \textit{Musas II} as well as a glimpse of blueish highlights in the mountain areas. Most of the samples contain low density of blue crystals. Figure~\ref{figure2}f (top) shows the sample with the bluest colour, dark violet, with medium density of blue crystals in layers 1 and 2. The average diameter of these crystals is approximately 2.0\,\,$\mu$m, see Fig.~\ref{figure5}c. The average size of the crystals indicates that Ferrario most likely worked with synthetic low--cost ultramarine rather than the expensive natural pigment.        
As expected, the Raman spectra for these crystals coincides with the spectra of both natural and artificial ultramarine pigments of the CHSOS, see Fig.~\ref{figure6}a. The mineral lazurite is an aluminosilicate of approximate formula (Na,Ca)$_{8}$(AlSiO$_{4}$)$_{6}$(SO$_{4}$,S,Cl)$_{2}$~\cite{osticioli2009analysis}, which is often associated with other silicate minerals like calcite (CaCO$_{3}$) and pirite (FeS$_{2}$)~\cite{rutherford1993artists}. The spectra of natural and synthetic reference pigments are not different enough to distinguish them because both have bands of the S$_{3}^{-}$ ion at 548 cm$^{-1}$ (symmetric stretching vibration), 258 cm$^{-1}$ (bending vibration) and 1096 cm$^{-1}$ (stretching vibration)~\cite{osticioli2009analysis}. What is interesting in this data is that our \textit{CrystalDistribution} software identified uniform particle distribution in both size and roundness, as reported in recent studies~\cite{osticioli2009analysis}, revealing the pigment to be artificial ultramarine with its smaller crystal size (0.5-5.0)\,$\mu$m~\cite{rutherford1993artists}.

\textit{Vermilion}

The purest and most saturated red, seen in the book in \textit{Musas II}, is likely vermilion red. The same pigment may have been washed or mixed with other colours to create the pinkish tones in the textiles.
In fact, red crystals were found in samples of light brown, pink and red areas. The redder the colour, the higher the density of crystals in the cross--section. For instance, in samples M1--54P and M2--69P, red crystals of approximately 1.4 \,$\mu$m and 3.0 \,$\mu$m, respectively, are found in low density in the paint layer, see Fig.~\ref{figure5}c. 
It is well--known that Italian schools commonly used a triad of vermilion, greenish--yellow and violet--blue to achieve chromatic equivalents~\cite{vanderpoel1901color}. Another reliable technique involving vermilion was to underpaint thin transparent touches of vermilion with another white pigment to achieve a pink tone for the rosy and ruddy portions of flesh and skin~\cite{mayer1975painter}. Ferrario likely used both of these techniques.
Raman spectra of red crystals were associated with natural and artificial vermilion pigments due to similar composition of mercury (II) sulfide, with characteristic signals at 251\,cm$^{-1}$ and 345\,cm$^{-1}$~\cite{petrova2019pigment} (see Fig.~\ref{figure6}a and Fig.~\ref{figure6}d). The particle size, fineness and uniformity of the vermilion crystals suggests the presence of artificial vermilion in these samples~\cite{rutherford1993artists}. 
It is interesting to note that even though vermilion itself does not fluoresce, the UVF picture of Fig.~\ref{figure2}d (top) shows slight pink fluorescence in the red mantle. This pigment can exhibit visible fluorescence induced by UV light due to the organic binder, if mixed with lead white or exposed to ammonia~\cite{delarie1982ultra}.

\textit{Chrome yellow}

 Yellow colours can be seen in elements like the harp in \textit{Musas I}, and ornaments including clothing details in both paintings. Yellow pigments washed or mixed with other colours is subtly perceptible in clouds, highlights, hair and skin tones. Chrome yellow was commonly used in artworks of the period.
 Yellow crystals at different density levels were found in all samples observed through OM. For instance, sample M1--75W has high density of yellow and orange crystals in the paint layer, see Fig.~\ref{figure4}b. We measured average crystal diameter of around 7.7 \,$\mu$m, see Fig.~\ref{figure5}c.
 Chrome yellow provided artists with a heavy saturated yellow pigment that balanced out other intense colours like red and blue. Ferrario may have chosen this pigment due to its reliability and popularity in the nineteenth and early twentieth centuries. The use of this pigment may have caused parts of the paintings to darken over time.
 Raman spectra show the characteristic bands of chrome yellow, also obtained for the CHSOS reference pigment. Chrome yellow consists of lead chromate with the following chemical formula: PbCrO$_{4}$~\cite{feller1986artist} along with Raman signal around 842\,cm$^{-1}$ caused by the Cr$^{(VI)}$--O stretching~\cite{maslar2001situ}, see Fig.~\ref{figure6}a and Fig.~\ref{figure6}d. A similar spectrum was obtained for the orange crystals observed in the samples. However, instead of the signal at 842\,cm$^{-1}$, a doublet is observed at 828 cm$^{-1}$ and 848\,cm$^{-1}$. Likewise, the observed signal at 361\,cm$^{-1}$ of the chrome yellow becomes four signals within roughly 337\,cm$^{-1}$. The latter is caused by the presence of oxidized lead (II) chromate, PbCrO$_{4}$.Pb(OH)$_{2}$, which is usually called chrome orange~\cite{castro2004micro}. We hypothesize that the presence of this compound is evidence of chrome yellow degradation, due to its sensitivity to light~\cite{feller1986artist}.
 Chrome yellow was introduced as a pigment in 1804~\cite{feller1986artist} and production increased in 1820~\cite{ward2008grove}. It was rather expensive in the first half of the nineteenth century~\cite{ball2008invention}, but the price had come down by the time Ferrario was working on \textit{Musas I} and \textit{Musas II}. Nowadays the pigment is seldom used due to its dangerous toxicity~\cite{feller1986artist}. 
 
\textit{Zinc and lead white}

White is a predominant colour in both \textit{Musas I} and \textit{Musas II}, most apparent in the clouds in both compositions. Lead white was the main white pigment on the market in Ferrario’s time, although zinc white was being developed as a safer alternative~\cite{eastaugh2008pigment}.
The fluorescent microscope revealed medium density of fluorescent particles in the first ground layer, see Fig.~\ref{figure4}c. However, in the optical microscope image, the fluorescent particles are indistinguishable from the rest of the components of the ground. Therefore, a white pigment with fluorescence characteristics~\cite{cosentino2014identification}, such as zinc white, must have been used in the preparation of the ground layer~\cite{christiansen2017artists}. The average diameter calculated for those particles was 5\,\,$\mu$m. 
The evidence that zinc white may be present includes its cold, flat tone~\cite{ball2008invention}, poor oil drying and low pigment density~\cite{eastaugh2008pigment}. Manufacturers began adding zinc white to other pigments as a lightening agent~\cite{feller1986artist}. Although zinc white was not a very popular pigment, a few artists were fond of it~\cite{ball2008invention}. Ferrario may have been one of them, but there is evidence that he might have used a mixture of zinc white with another white, or a pigment mixed with zinc white. 
Micro-Raman analysis was performed on these fluorescent particles. The spectra obtained allowed us to identify zinc white, comparing it to the CHSOS reference pigment. The Raman spectra correspond to bands at approximately 447\,cm$^{-1}$ and 610\,cm$^{-1}$~\cite{cardeira2016analytical} which agrees with the fluorescence microscopy because ZnO fluoresces when exposed to ultraviolet light~\cite{christiansen2017artists}.
As mentioned above, the second ground layer is homogeneous white, without coloured pigments. Measurements using SEM--EDX taken on a specific region reveal the following percentages: 27$\%$ of carbon, 14$\%$ of oxygen, 2$\%$ of calcium, 53$\%$ of lead and 4$\%$ of zinc. This significant amount of lead, carbon and oxygen could indicate the presence of lead white pigment (PbCO$_{3}$)$_{2}$·Pb(OH)$_{2}$~\cite{Fitz1997artist}. We recommend additional studies to corroborate this hypothesis. 
Zinc white was developed to reduce the incidence of lead poisoning~\cite{feller1986artist, ball2008invention}. However, at 8 French francs per pound, it was four times the cost of lead white\cite{feller1986artist}, so in many cases artists were slow to adopt the newer product.

      \item \subsubsection*{Mapping of damage} 

Thirdly, intensity analysis can be a first step towards identifying deterioration agents and selecting possible restoration techniques. Multispectral Imaging (MSI), combined with \textit{RegionsOfInterest}, shows the type and severity of damage on artwork, allowing us to select areas of interest with visual atypical behaviours. The main observations are shown in Fig.~\ref{figure7}. Panels Fig.~\ref{figure7}a and Fig.~\ref{figure7}b show the paintings themselves. Panels Fig.~\ref{figure7}a.6 and Fig.~\ref{figure7}b.6, show a schematic of the paintings for easier reference.
We identified ten categories of damage: 1) moisture marks Fig.~\ref{figure7}b.2, 2) detachment of canvas (bubbles) Fig.~\ref{figure7}b.1, 3) cracks Fig.~\ref{figure7}b.3, 4) craquelure, Fig.~\ref{figure7}a.3, 5) loss of paint and exposed canvas Fig.~\ref{figure7}a.1, 6) human fingerprints Fig.~\ref{figure7}b.4 (see supplementary information Fig.~\textcolor{blue}{S1}), 7) spots Fig.~\ref{figure7}a.2, 8) holes Fig.~\ref{figure7}a.5, 9) insect specks and perforation Fig.~\ref{figure7}b.5 and 10) fungi Fig.~\ref{figure7}a.4. 

Moisture marks are among the most prevalent types of damage~\cite{Mecklenburg2007} caused by leaks, dampness or floods which can weaken and detaches the adhesive layer bonding the canvas to the wood support~\cite{Mecklenburg2007}. Possible effects include bulges or air bubbles as observed in Fig.~\ref{figure7}b.1) and even cracks, Fig.~\ref{figure7}b.3). The glue binding the ground to the canvas may detach in relative humidity above 80\,\%. As the canvas shrinks paint layers may separate, leaving the canvas exposed, Fig.~\ref{figure7}a.1.
Craquelure may also cause paint loss~\cite{bucklow1997}. This cracking in the paint layer reflects the materials and technique used by the artist and its pattern can even provide information on the creative process~\cite{bucklow1997}. The location of the craquelure on \textit{Musas I} and \textit{Musas II} suggests it was caused by moisture inside the paintings.

Three predominant types of deterioration on \textit{Musas I} and \textit{Musas II} are: moisture marks Fig.~\ref{figure7}b.2, craquelure Fig.~\ref{figure7}a.3, and loss of paint and exposed canvas Fig.~\ref{figure7}a.1. We also observed human--caused damage, such as fingerprints Fig.~\ref{figure7}b.4 (see supplementary information Fig.~\textcolor{blue}{S1}), deep holes Fig.~\ref{figure7}a.5 and spots Fig.~\ref{figure7}a.2. Finally, with the help of software and multispectral images, we identified regions with insect specks, perforation and fungi colonization.
Temperature and humidity cycles cause mechanical stress that may result in structural damage, such as fracturing and flaking~\cite{brezoczki2017study}. While assessing the risk of biodeterioration in a tropical climate, Bhattacharyya \textit{et al.} found that higher relative humidity and temperature leads to higher fungal load~\cite{Bhattacharyya2016}. The data collected at the NTCR, see Fig.~\ref{figure8}, revealed temperatures between 22.7\,$^{\circ}$C and 25.1\,$^{\circ}$C, with an average of 23.3\,$^{\circ}$C, see Fig.~\ref{figure8}d. Meanwhile, relative humidity ranged from 60.2\,\% to 78.5\,\%, with an average of 70.3\,\%, see Fig.~\ref{figure8}e. These conditions are considered very high--risk for biodeterioration. Air pollution and radiation levels are also known to contribute to the deterioration process~\cite{smielowska2017indoor,elgohary2016monitoring,ajmat2011lighting,mohelnikova2018analysis}; ergo, we recommend monitoring these parameters in the future.
Fungi can cause severe aesthetic and structural deterioration, since composite materials provide a number of substrates for microbial growth~\cite{lopez2013}. Most fungi penetrate the fiber lumen where they grow a mycelium \cite{szostak2004biodeterioration,mazzoli2018back}, and some have been found to secrete cellulolytic enzymes that dissolve cellulose fibers, causing loss of strength, elasticity, structure, and ultimately damaging the canvas~\cite{poyatos2018physiology}. Additionally, a number of fungi and bacteria are known to produce pigments or discoloration on artwork ~\cite{sterflinger2013microbial,gutarowska2017historical}. Identifying possible areas of biodeterioration with MSI--coupled software tools in a great advantage, since fungal proliferation is often only noted once it has caused significant damage~\cite{bhattacharyya2016biodeterioration}.

\end{enumerate}

\section*{Conclusions}
This research aimed to provide a baseline diagnostic of the conservation state of the paintings \textit{Musas I} and \textit{Musas II} using comprehensive and multidisciplinary methodologies. In particular, this multi--analytical study (MAS) focused on the challenges of painting conservation in a tropical climate, and sought to provide fundamental knowledge to prevent and slow down this decay. Novel software allowed us to gather data in a non--destructive manner. One of the most significant outcomes of this study, made possible by our \textit{RegionsOfInterest} software, was the identification of areas of particular interest and assessment of the artist's colour palette through minimally invasive sampling. We found that zinc white, lead white, chrome yellow, lead red, viridian, vermilion, and ultramarine blue pigments make up approximately 50\% of the total pigment composition in the artwork, demonstrating one of the ways this innovative technology can be applied to conservation, restoration and even art history research on large--format paintings. Our second major development was the \textit{CrystalDistribution} software, which scans sample cross--sections and calculates pigment crystal number and size distributions. Using this tool, we were able to identify synthetic vermilion and ultramarine blue in the paintings, as opposed to natural pigments. This finding has important implications in developing affordable, efficient, accessible, and innovative non--destructive laboratory analyses, especially given the current limited access to cultural heritage sites. The third significant finding of this study was the ability to map damaged areas on the works, using \textit{RegionsOfInterest} to guide future restoration efforts. The results in this study have demonstrated the importance of a MAS approach to large--format paintings and the value of our two novel software tools for art conservation studies. In further research, we believe that further software tools could be developed to reveal the artist’s intentions and creative process, for example, using machine learning techniques similar to deep neural networks~\cite{Fiorucci2020}.

\section*{Materials and Experimental}

\subsection*{Multispectral Imaging: photography acquisition}

We generated visible (VIS), infrared (IR), infrared false colour (IRFC) and ultraviolet (UV) photographs. VIS photos show possible superficial damage and suggest the colours used by the artist, which may have changed with age or restoration \cite{hayem2015characterizing}. IR imaging shows background layers; such as underlying sketches, drawings and corrections by the artist \cite{pronti2019post,comelli2008portable}. UV imaging shows deeper damage and compounds that fluoresce when excited with UV radiation, like certain pigments, binders and glues~\cite{comelli2008portable,pelagotti2008multispectral}.

Photography acquisition was based on the procedure described by Cosentino (2014)~\cite{cosentino2014identification}. Due to the large dimensions of the paintings (2.97\,m (w), 6.17\,m (h)), a set of 30 pictures was obtained for each spectral region: VIS, IR, ultraviolet fluorescence (UVF) and ultraviolet reflectance (UVR). Since the artworks are located on the ceiling, the photographic equipment was placed on a grid marked on the floor roughly 3.0\,m below. Both sets of pictures were obtained using a modified camera (Nikon D7200) with the following filters: XNiteCC1 M52 (VIS), IR (IR), XNiteCC1 M5 and B+W 52\,mm 403 (Ultraviolet fluorescence), XNiteCC1 M5 and Baader UV/IR Cut/L-Filter 2” RISE(UK) 52\,mm-48\,mm (Ultraviolet reflected). Camera settings were adjusted for high quality photographs (24 megapixels)~\cite{pronti2019post,comelli2008portable,pelagotti2008multispectral,cosentino2014identification,cosentino2013practical}. The camera’s experimental conditions were: F5, ISO 100 and exposition time of $\frac{1}{4}$\,s (for VIS and IR), F4, ISO 100 and exposition time of 4\,s (for UV)~\cite{pronti2019post,comelli2008portable,pelagotti2008multispectral,cosentino2014identification,cosentino2013practical}. 

Two 500\,W halogen lamps were used for visible and infrared photography (situated approximately 2.0\,m under the paintings) and four fluorescent lamps were used for ultraviolet photography (situated approximately 1.5\,m under the paintings). The white reference in the RMI Conservation Target (2017 Robin Myers Imaging®)~\cite{RMI_2020} used for the white balance was placed on one border of the painting. We used Cultural Heritage Open Source Pigments Checker (CHSOS) pigment reference for the photos in every region; and Lightroom Classic® and Photoshop® for lens corrections and generation of Infrared False colour (IRFC) pictures, respectively. We generated panoramic images in each spectral region using PTGUI® Software. For IRFC we used Adobe Photoshop® to overlay VIS and IR images. Both must align precisely to exchange their red, green and blue channels as follows: in VIS the blue channel was suppressed and substituted with the green one, then the red channel substituted the green channel, and the IR red replaced the VIS ex--red channel~\cite{hayem2015characterizing,pronti2019post}. 

\subsection*{Software development}

Given the importance of digital image processing in painting analysis\cite{barni2005image , bradski2008learning}, we developed two novel software tools, \textit{RegionOfInterest} and \textit{CrystalDistribution}, written in Python$3$ using the OpenCV library~\cite{culjak2012brief, opencv2008computer} for pixel colour management in the HSV (Hue-Saturation-Value) colour space~\cite{agoston2005computer}. Both are available for free at the GitHub repository https://github.com/andress5990/16 \_Intensity\_Analysis and are open to modification and improvements.

\subsubsection*{RegionOfInterest program for luminosity analysis of paintings}

\textit{RegionOfInterest} has two stages to quantify pixel intensity value. The first takes a user--created, hand--drawn image region of any shape and size. The second separates all pixels in the image with an intensity value within a user--selected range based on the stage one results. These two execution routines use the OpenCV colour to gray transformation and classify pixels in a range from 0 (black or null intensity) to 255 (white or full intensity)~\cite{minichino2015learning , agoston2005computer}. The program then generates histograms of "Count vs intensity", according to the distribution of the intensity values. The second stage generates a colour image containing only pixels with intensity values within the selected range, showing their location on the original image. Intensity measurement uses the OpenCV $cv2.colour\_BGR2GRAY$ library function \cite{minichino2015learning , bradski2008learning}. It was compared with BT.601 and BT.709 colour to gray transformation methods~\cite{gunecs2016optimizing, kylander1999gimp}, with uncertainties of $\pm$0.2\,\% and $\pm$2.1\,\% respectively. Program execution time was approximately 1\,s. It is important to note that processing time depends on image format and size and computer processing power. These analyses are also heavily dependent on image quality and lighting conditions. 

\subsubsection*{CrystalDistribution software tool to measure pigment crystals}
\textit{CrystalDistribution} identifies the number, diameter and area of crystals of a selected colour in 50x magnified images. A code change is required to use this software on other magnifications; this particular feature is not user--selectable. The code has three functions: (1) image analysis, (2) numerical calculations, and (3) crystal counting. The first consists of slicing a microscopic image. User made cuts allow both specific region and layer by layer analyses. We established hue, saturation and component ranges to create a mask for each colour layer. Each mask is applied to the cuts and saved as an image.
We then calculate the area of the crystals by applying a mask of the corresponding colour to identify their contour, and finding their center. This information is used to generate "Crystal count vs diameter" histograms (in ranges of 2\,$\mu$m from 0\,$\mu$m to 25\,$\mu$ m).

\subsection*{Sampling along with microscopy and spectroscopy analysis}

Under the supervision of the Manager of the NTCR Conservation Department we collected 15--millimetric samples (0.5\,mm$^2$ - 3\,mm$^2$) with a scalpel from already--damaged regions, so as to minimize risk of further damage. Table~\ref{table1} presents an overview of the main characteristics of these samples. We used a scaffold, since the paintings are located on the ceiling. Figure~\ref{figure2}f (top) and Figure~\ref{figure2}f (bottom) show the grid for sample collection and labeling. To set up the samples, cross--sections were embedded in epoxy resin (Fibrocentro®) and polished with water sanding sheets (3M®) of 9.2\,$\mu$m, 6.5\,$\mu$m 3\,$\mu$m and 1\,$\mu$m grit. Samples were sanded for approximately 3 minutes with each sheet, in descending order. We then used a polisher (EcoMet 30, Buehler®) for two 10--minute cycles with 15\,$\mu$m and 3\,$\mu$m alumina abrasive (Allied High Tech Products®), and the following conditions: 150\,rpm for stage and 150\,rpm and 40\,N force for the support arm, which rotated opposite to the stage. Cross--sections were examined by Optical Microscope, SEM--EDX and Raman.

\subsubsection*{Optical microscopy}

 Cross--sections were examined with optical microscopy at 5x, 10x, 20x and 50x magnifications, using a Nikon Eclipse LV100ND light microscope equipped with a Nikon DS-fi3 digital camera. Microscopic images with reflected light reveal the paintings' stratigraphy and the colour of individual pigment grains~\cite{mahmoud_2014}. 10x magnification showed that cross--sections were composed typically of 2 to 4 layers, see Fig.~\ref{figure4}b. At 20x and 50x magnifications we observed pigment crystals in the painting layer ranging from about 3\,$\mu$m to 18\,$\mu$m, and fossils of 20\,$\mu$m to 50\,$\mu$m in the ground layer. Fluorescent crystals in the ground layer were observed through Leica DMi8 Invert fluorescence microscope and images were capture at 5X, 10X and 100X. The signal was measured with a \textit{DAPI LP} filter, with a range of (420 $\pm$ 20)\,nm for the excitation band pass filter and a range of (457 $\pm$ 20)\,nm for the suppression band pass filter.
 
\subsubsection*{Scanning Electron Microscope – Energy Dispersive X Ray Spectroscopy (SEM--EDX)}

We measured the paint stratigraphy of 8 samples with EDX to determine elemental composition of the layers. Mapping and single point analysis was conducted using a scanning electron microscope HITACHI S--3700n coupled with an IXRF Systems Detector Energy--Dispersive X--ray spectrometer. The samples were uncoated with gold and analysed in low vacuum conditions and backscattering electron mode, under the following analytical conditions: 15\,kV accelerating voltage, 80\,$\mu$A beam current, working distance ranging from about 5.5\,mm to 10.9\,mm and collection time for mapping measurements was typically 30\,min. 

\subsubsection*{Fourier Transform Infrared - Attenuated Total Reflectance Spectroscopy (FTIR--ATR)}

We used FTIR--ATR to identify the chemical composition of the ground layer. Three samples (see Table~\ref{table1}) not embedded in resin were analyzed by placing their inner layer in direct contact with the ATR crystal. We used a spectrometer (PerkinElmer Frontier®) equipped with an ATR detector. Spectra were collected in transmittance mode, with 16 scans, in the wavenumber range from 4000\,cm$^{-1}$ to 650\,cm$^{-1}$, at 1\,cm$^{-1}$ spectral resolution. 

\subsubsection*{Micro--Raman Spectroscopy}

We used micro--Raman Spectroscopy to identify the pigments in 16 samples, using a WiTec alpha 300R micro--Raman Spectrometer with a diode laser of 532\,nm and operating power ranging from 0.06\,mW to 0.18\,mW. Sample irradiation diameter of the laser was approximately 1\,$\mu$m. Spectra were measured from 0\,cm$^{-1}$ to 3000\,cm$^{-1}$. Measurement conditions to provided a low signal/noise ratio were: 100x magnification, 50 cycles of accumulation and 0.5\,s of integration time. Spectra were compared with reference pigments from CHSOS.

\subsection*{Analysis of the environmental and biological factors}
MSI in VIS and UV, coupled with our computational tools, identified regions of possible biodeterioration. These regions were sampled for fungi and bacteria. Samples were collected using sterile cotton swabs from quadrants 54, 59 and 60 on \textit{Musas I} and 50, 56, 59 and 69 on \textit{Musas II}. They were transported in Phosphate Buffered Saline~\cite{Camuffo2014heritage} and inoculated for culture in Potato Dextrose Agar (PDA) medium around 23\,$^{\circ}$C. Over the next three months, fungi were isolated and observed through optical microscopy at 100X in lactophenol cotton blue and Gram stain. In order to measure periodically environmental variables at the NTCR, six portable stations were developed with a lithium-ion polymer (LiPo) battery, and an Internet of Things (IoT) microcontroller device. The sensors used are the following: SHTC3 (Sensirion) to measure temperature and relative humidity, SGP30 (Adafruit) breakout board was used to monitor the variation of CO$_{2}$ concentration, and finally TSL2591 (Adafruit) breakout board was used to measure illuminance levels.  

\bibliography{sample}

\section*{Acknowledgements}

We would like to thank Jocelyne Alcántara--García from University of Delaware for her careful reading of the manuscript and helpful comments and suggestions during the development of this research. We give special thanks to Tiffany Fernández Estrada from the School of Arts at the University of Costa Rica for providing us with key insight on the applications of our software. We also would like to thank the Geochemistry Laboratory from the Central American School of Geology and the Inorganic Chemistry Section from the Chemistry Department at the University of Costa Rica for all the valuable assistance in stratigraphy sample preparation and polishing. We want to acknowledge the support of Marcial Garbanzo Salas in the design and construction of environmental sensors used to detect temperature and humidity variation. We wish to thank Marielos Mora López and Fernando Morales Calvo for running preliminary microbiological tests on the paintings. A special thanks to Jennifer Tucker for reviewing the manuscript and improving the English writing. We are grateful for the support given by the Vicerrectoría de Investigación at the Universidad de Costa Rica to carry out this research work.

\section*{Author contributions statement}

M.D.B.M. designed methodology and carried out the majority of the experiments. T.Z.S. produced the artwork shown in Fig.~\ref{figure1}. R.E.A.T., J.M.V., K.A.U., M.G.T. and F.L.F. assisted in experiments and characterizations. A.C.S. and J.S.S. programmed and developed the computational tools. K.W.Q., T.Z.S. and C.M.C. developed the historical context of paintings and pigments studied. K.W.Q., M.D.B.M. and T.Z.S. prepared the final version of all figures. O.A.H.S. conceived and led the project, was involved in experimental design, data analysis and interpretation. M.D.B.M., T.Z.S., R.E.A.T. and O.A.H.S. wrote the paper and all co--authors discussed the results, and commented on the manuscript. All authors have given approval to the final version of the manuscript.

\section*{Additional information}

\begin{figure}
    \centering
    \includegraphics[scale=0.12]{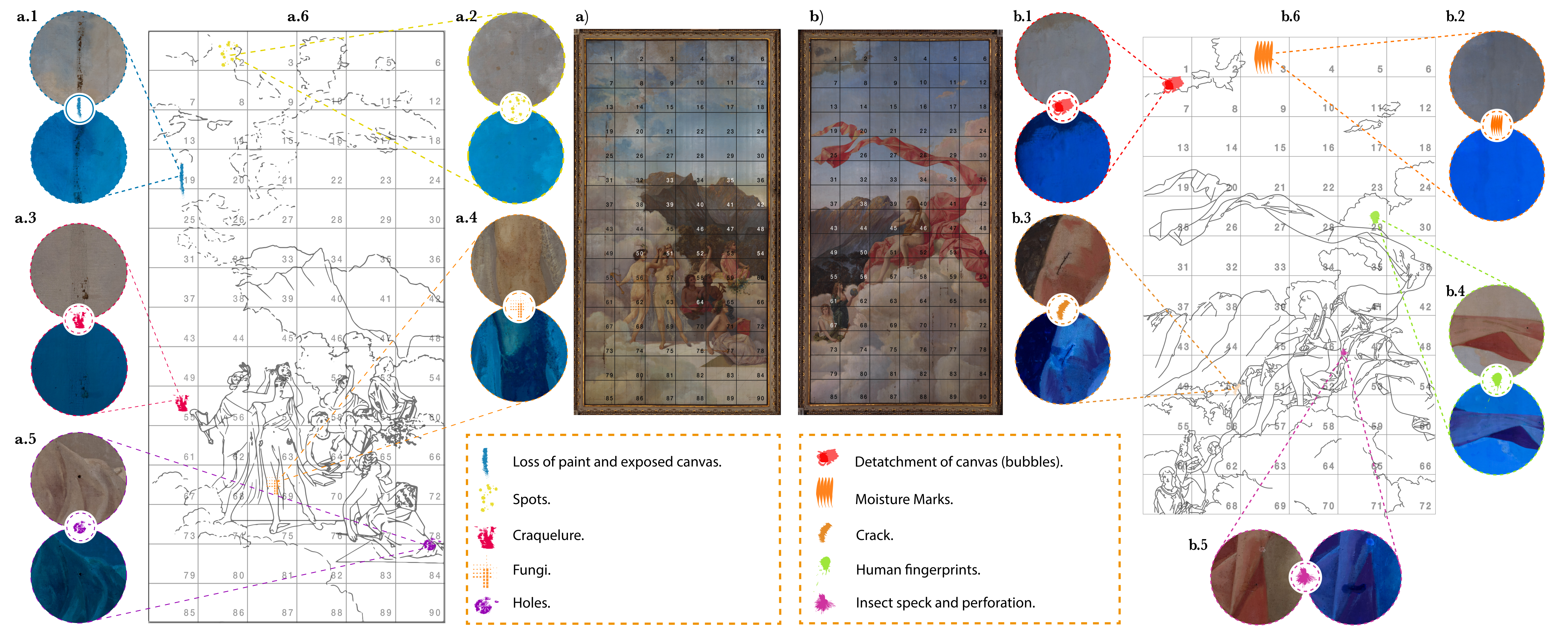}
    \caption{\textbf{Traces of time: Categorising damage observed in the paintings.} Visible photographs of \textbf{(a)} \textit{Musas II} and \textbf{(b)} \textit{Musas I} with the working grid. \textbf{(a.6)} \textit{Musas II} and \textbf{(b.6)} \textit{Musas I} diagrams showing locations of the ten most significant types of damage, with visible and UVF images of each: \textbf{a.1.)} Loss of paint and exposed canvas. \textbf{a.2.)} Spots. \textbf{a.3.)} Craquelure. \textbf{a.4.)} Fungi. \textbf{a.5.)} Holes. \textbf{b.1.)} Detachment of canvas (bubbles). \textbf{b.2.)} Moisture marks. \textbf{b.3.)} Cracks. \textbf{b.4.)} Human fingerprints. \textbf{b.5.)} Insect specks and perforation.}
    \label{figure7}
\end{figure}

\begin{figure}
    \centering
   \includegraphics[scale=0.3]{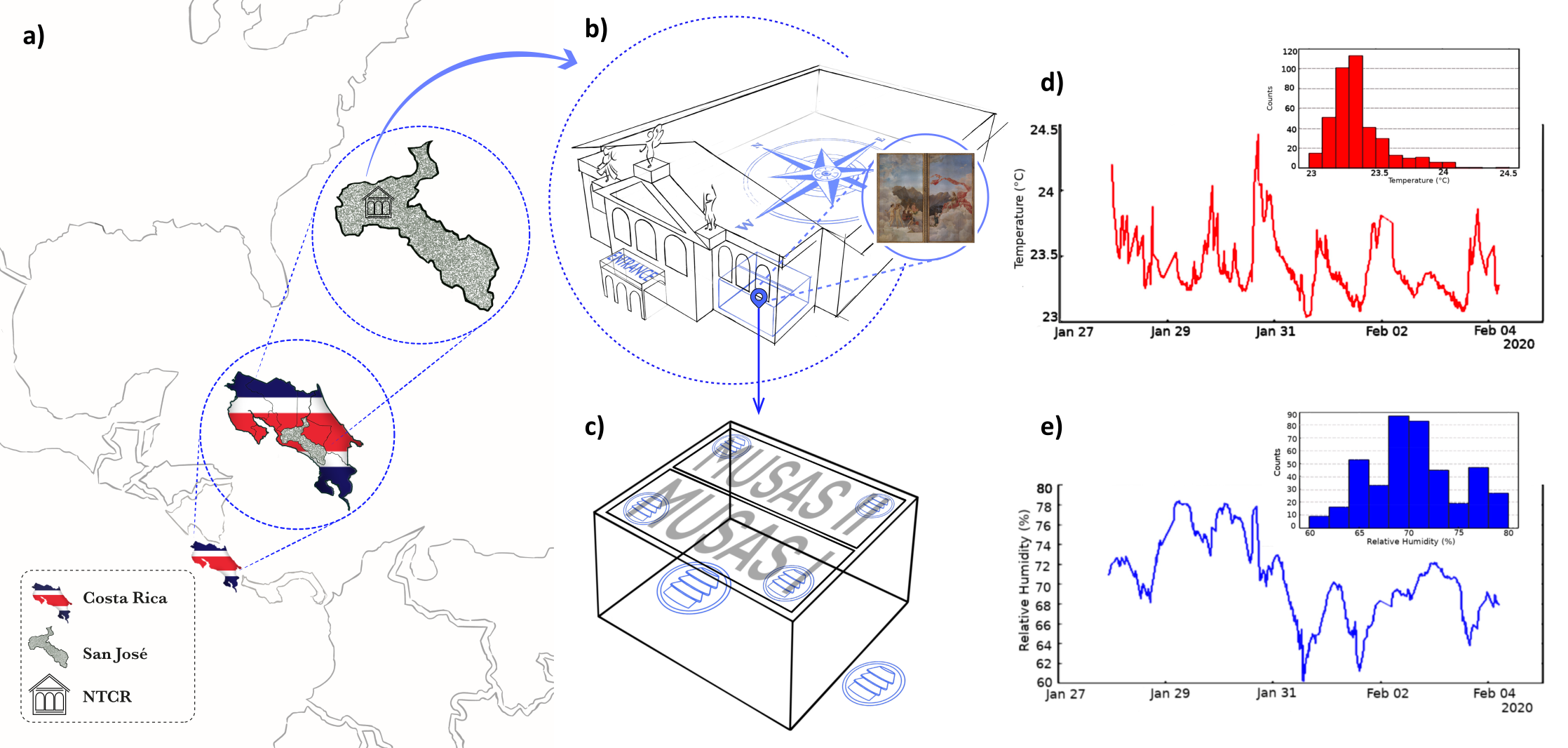}
   \caption{\textbf{Conservation in the tropics: Monitoring the paintings' environment.} \textbf{(a)} National Theatre of Costa Rica in San José (NTCR), Costa Rica, Central America. \textbf{(b)} \textit{Musas I} and \textit{Musas II} are located on the ceiling, at a height of about 3.5\,m in the ladies’ lounge on the ground floor of the Theatre. \textbf{(c)} Location of the environmental sensors. \textbf{(d)} Temperature and \textbf{(e)} humidity variations in the paintings' microclimate.}
    \label{figure8}
\end{figure}

\begin{table}[H]
\centering
\caption{\textbf{Paint stratigraphy samples}. Pigment density, observed through optical microscopy, is categorized as l = low, m = medium or h = high, for each type of pigment: L = lead red, Vi = viridian, U = ultramarine, Ve = vermilion and C = chrome yellow. The average diameter of the crystals, in micrometers, is indicated in parenthesis next to the pigment symbols. Analysis symbology: OM = optical microscopy, R = Raman spectroscopy, EDX = energy dispersive X ray spectroscopy, UV = fluorescence microscopy and FTIR--ATR = Fourier transform infrared - attenuated total reflectance spectroscopy. For FTIR--ATR analysis, samples were not embedded. The rest of the analyses were carried out on cross--sections of samples.}
%\label{tab:my-table}
\begin{tabular}{|c|c|c|c|c|c|c|}
\hline
\textbf{Painting} & \textbf{\begin{tabular}[c]{@{}c@{}}Sample\\ Name\end{tabular}} & \textbf{\begin{tabular}[c]{@{}c@{}}Visible \\ colour\end{tabular}} & \textbf{\begin{tabular}[c]{@{}c@{}}Grid \\ location\end{tabular}} & \textbf{Pigments density} & \textbf{\begin{tabular}[c]{@{}c@{}}Crystal average   \\ diameter ($\mu m$)\end{tabular}} & \textbf{Analysis} \\ \hline
\multirow{9}{*}{\textbf{Musas I}} & \multirow{2}{*}{M1-45B} & \multirow{2}{*}{Dark   blue} & \multirow{2}{*}{45} & Layer 1: Ve(h), U(m), C(m) & \multirow{2}{*}{Ve(4.7), U(2.0), C(1.2)} & \multirow{2}{*}{MO, R, EDX} \\
 &  &  &  & Layer   2: C(m), U(m) &  &  \\ \cline{2-7} 
 & M1-54P & Pink & 54 & Layer 1: Ve(m),   L(l) & Ve(1.4), L(21) & MO, R \\ \cline{2-7} 
 & M1-54R & Red & 54 & Layer   1: Ve(h), C(m) &  & MO, R, SEM, UV \\ \cline{2-7} 
 & M1-58W & White & 58 & Layer   1: C(l) &  & MO, R, EDX \\ \cline{2-7} 
 & \multirow{2}{*}{M1-60W} & \multirow{2}{*}{White} & \multirow{2}{*}{60} & Layer 1: C(l) &  & \multirow{2}{*}{MO, R} \\
 &  &  &  & Layer   2: U(l), C(l) &  &  \\ \cline{2-7} 
 & \multirow{2}{*}{M1-75W} & \multirow{2}{*}{White} & \multirow{2}{*}{75} & Layer 1: Ve(l),   C(l) & \multirow{2}{*}{C(7.7), U(3.4)} & \multirow{2}{*}{\begin{tabular}[c]{@{}c@{}}MO, R, SEM, EDX, \\ FTIR-ATR, UV\end{tabular}} \\
 &  &  &  & Layer   2: C(h), U(m) &  &  \\ \hline
\multirow{14}{*}{\textbf{Musas II}} & \multirow{2}{*}{M2-27B} & \multirow{2}{*}{Light   blue} & \multirow{2}{*}{27} & Layer 1: U(m) & \multirow{2}{*}{} & \multirow{2}{*}{MO, R, UV} \\
 &  &  &  & Layer   2: U(m), C(m) &  &  \\ \cline{2-7} 
 & \multirow{2}{*}{M2-31W} & \multirow{2}{*}{White} & \multirow{2}{*}{31} & Layer 1: Ve(l) & \multirow{2}{*}{} & \multirow{2}{*}{\begin{tabular}[c]{@{}c@{}}MO, R, EDX,\\ FTIR-ATR, UV\end{tabular}} \\
 &  &  &  & Layer   2: U(h), C(h) &  &  \\ \cline{2-7} 
 & \multirow{2}{*}{M2-33Br} & \multirow{2}{*}{Brown} & \multirow{2}{*}{33} & Layer 1: Ve(m), U(l), C(l) & \multirow{2}{*}{} & \multirow{2}{*}{MO, R, EDX, UV} \\
 &  &  &  & Layer   2: U(m), C(m) &  &  \\ \cline{2-7} 
 & M2-39B & Dark   blue & 39 & Layer 1: U(m),   C(m), Ve(l) & U(4.7),   C(1.4), Ve(2.3) & MO, R, EDX, UV \\ \cline{2-7} 
 & M2-63W & White & 63 & Layer   1: C(h), U(h) &  & R \\ \cline{2-7} 
 & M2-69P & Pink & 69 & \begin{tabular}[c]{@{}c@{}}Layer   1: Vi(l), Ve(l),\\ L(l), U(l), C(l)\end{tabular} & \begin{tabular}[c]{@{}c@{}}Vi(5.1),   Ve(3.0), \\ L(0.6), U(2.5), C(0.6)\end{tabular} & MO, R \\ \cline{2-7} 
 & \multirow{2}{*}{M2-70L} & \multirow{2}{*}{Light   purple} & \multirow{2}{*}{70} & Layer 1: Ve(m), U(m), C(l) & \multirow{2}{*}{} & \multirow{2}{*}{R} \\
 &  &  &  & Layer   2: C(h), U(h) &  &  \\ \cline{2-7} 
 & \multirow{2}{*}{M2-73Br} & \multirow{2}{*}{Brown} & \multirow{2}{*}{73} & Layer 1: U(m), C(m) & \multirow{2}{*}{C(4.8),   U(4.13)} & \multirow{2}{*}{MO, R} \\
 &  &  &  & Layer   2: U(m), C(m) &  &  \\ \cline{2-7} 
 & M2-83Br & Brown & 83 &  &  & FTIR-ATR \\ \hline
\end{tabular}
\label{table1}
\end{table}

\end{document}